# Time Fractional Formalism: Classical and Quantum Phenomena

Hosein Nasrolahpour[*]

## Abstract

In this review, we present some fundamental classical and quantum phenomena in view of time fractional formalism. Time fractional formalism is a very useful tool in describing systems with memory and delay. We hope that this study can provide a deeper understanding of the physical interpretations of fractional derivative.

*This work is dedicated to the soul of my father*

## 1- Introduction

Fractional calculus is a very useful tool in describing the evolution of systems with memory, which typically are dissipative and to complex systems. Complex systems include very broad and general class of systems and materials. For instance, glasses, biopolymers, biological cells, porous materials, amorphous semiconductors and liquid crystals can be considered as complex systems. Scaling laws and self-similar behavior are supposed to be fundamental features of complex systems. In recent decades the fractional calculus and in particular the fractional differential equations has attracted interest of researches in several areas including mathematics, physics, chemistry, biology, engineering and economics [1-4].Applications of fractional calculus in the field of physics have gained considerable popularity and many important results were obtained during the last years. Some of the areas of these applications include: classical mechanics [8-11], classical electromagnetism [32-38], special relativity [39, 40], non-relativistic quantum mechanics [43-50] and relativistic quantum mechanics and field theory [51-58]. Despite these various applications, there are some important challenges. For example physical interpretation for the fractional derivative is not completely clarified yet. In this review, we aim to present some aspects of physical interpretation for the fractional derivative by studying the behavior of fundamental classical and quantum phenomena within the framework of time fractional formalism. In the following, fractional calculus is briefly reviewed in Sec. 2. The fractional relaxation and oscillation process are discussed in Sec. 3. Time fractional Maxwell's equations are presented in sec. 4. In Sec. 5 time fractional Schrödinger equation and time fractional Pauli equation are given. Finally in Sec. 6, we will present our summary and discussion.

## 2- Mathematical tools: Fractional calculus

Although the application of Fractional calculus has attracted interest of researches in recent decades, it has a long history when the derivative of order $0.5$ has been described by Leibniz in a letter to L'Hospital in 1695. Fractional calculus is the calculus of derivatives and integrals with arbitrary (real or even complex) order, which unify and generalize the notions of integer order differentiation and n-fold integration, which have found many applications in recent studies to model a variety processes from classical to quantum physics. In the following, we briefly revisit essentials of fractional calculus.

### 2.1. The Caputo fractional derivative operator

The commonest way to obtain a fractional differential equation for describing the evolution of a typical system is to generalize the ordinary derivative in the standard differential equation into the fractional derivative. Fractional differential equation can be include for instance derivative of order $0.5, \sqrt{2}, \pi$ and so on. Since the age of Leibniz various types of fractional derivatives have been proposed. In fact, the definition of the fractional order derivative is not unique and there exists several definitions including, Grünwald–Letnikov, Riemann-Liouville, Weyl, Riesz and Caputo for fractional order derivative. Fractional differential equations defined in terms of Caputo derivatives require standard boundary (initial) conditions. Also the Caputo fractional derivative satisfies the relevant property of being zero when applied to a constant. For these reasons, in this paper we prefer to use the Caputo fractional derivative. The left (forward) Caputo fractional derivative of a time dependent function $f(t)$ is defined by

$$_0^c D_t^\alpha f(t) = \frac{1}{\Gamma(n-\alpha)} \int_0^t (t-\tau)^{n-\alpha-1} f^{(n)}(\tau) d\tau \qquad \alpha > 0, t > 0 \qquad (1)$$

---

[*] Correspondence: Hosein Nasrolahpour, E-mail: hnasrolahpour@gmail.com

Where, n is an integer number and $\alpha$ is the order of the derivative such that n-1< $\alpha$ <n and $f^{(n)}(\tau)$ denotes the n-th derivative of the function $f(\tau)$. For example when $\alpha$ is between 0 and 1, we have

$$_0^c D_t^\alpha f(t) = \frac{1}{\Gamma(1-\alpha)} \int_0^t (t-\tau)^{-\alpha} \frac{\partial f(\tau)}{\partial \tau} d\tau \qquad 0 < \alpha < 1 \qquad (2)$$

As we can see from the above equations Caputo derivative implies a memory effects by means of a convolution between the integer order derivative and a power of time. Also the Laplace transform to Caputo's fractional derivative gives

$$L\{_0^c D_t^\alpha f(t)\} = s^\alpha F(s) - \sum_{m=0}^{n-1} s^{\alpha-m-1} f^{(m)}(0) \qquad (3)$$

where, $F(s)$ is the Laplace transform of $f(t)$.

**2.2. The Mittag-Leffler function**

During the recent years the Mittag-Leffler (ML) function has caused extensive interest among physicist due to its role played in describing realistic physical systems with memory and delay. It was originally introduced by G.M. Mittag-Leffler in 1902[5]. The ML function is such a one-parameter function defined by the series expansion as

$$E_\alpha(z) = \sum_{k=0}^\infty \frac{z^k}{\Gamma(1+\alpha k)} \qquad z \in \mathbf{C}, \alpha > 0 \qquad (4)$$

And its general two-parameter representations is defined as

$$E_{\alpha,\beta}(z) = \sum_{k=0}^\infty \frac{z^k}{\Gamma(\beta+\alpha k)} \qquad z \in \mathbf{C}, \beta \in \mathbf{C}, \alpha > 0 \qquad (5)$$

where **C** is the set of complex numbers and $\Gamma(\alpha)$ denotes the Gamma function. This function is in fact a generalization of the exponential function. For example, for the special case of $\alpha = 1$, the ML function Eq. (4) reduces to the exponential function $E_1(z) = e^z$.

Furthermore, since the ML function generalizes the exponential function, the Euler identity for an exponential function with a complex argument (i.e., $e^{i\theta} = \cos(\theta) + i\sin(\theta)$) can also be written for the ML function in a similar manner. So we have

$$E_\alpha(i\theta) = \cos_\alpha(\theta) + i\sin_\alpha(\theta) \qquad (6)$$

Where $\sin_\alpha(\theta)$ and $\cos_\alpha(\theta)$ are sine and cosine ML functions respectively and defined as

$$\sin_\alpha(\theta) = \sum_{n=0}^{\infty} \frac{(-1)^n (\theta)^{2n+1}}{\Gamma((2n+1)\alpha+1)}, \qquad \cos_\alpha(\theta) = \sum_{n=0}^{\infty} \frac{(-1)^n (\theta)^{2n}}{\Gamma(2n\alpha+1)} \qquad (7)$$

Also, it is notable that although exponential function possesses the semigroup property (i.e., $e^{a(z_1+z_2)} = e^{az_1} e^{az_2}$) the function $E_\alpha(az^\alpha)$ does not possess the semigroup property in general [6] (this property leads to important results in fractional quantum mechanics [48]). Mittag-Leffler function, as a generalized exponential function, naturally arises in the solutions of ordinary differential equations of arbitrary (non-integer) order. Therefore the Laplace transform for ML function will be very useful in solving fractional differential equations:

$$L\{t^{\alpha m+\beta-1} E_{\alpha,\beta}^{(m)}(\pm \lambda t^\alpha)\} = \frac{m!\, s^{\alpha-\beta}}{(s^\alpha \mp \lambda)^{m+1}} \qquad (8)$$

Where $s > |\lambda|^{\frac{1}{\alpha}}$.

**3- Classical mechanics: fractional relaxation and fractional oscillation**

The fundamental processes in physics are described by equations for the time evolution of a quantity $X(t)$ in the form:

$$\frac{dX(t)}{dt} = -LX(t) \qquad (9)$$

where $L$ can be both linear or nonlinear operator. For instance there are many relaxation phenomena in nature whose relaxation function obeys the simple approximate equation

$$\tau \frac{dx(t)}{dt} + x(t) = 0 \qquad (10)$$

We can write the above equation as

$$\frac{dx(t)}{dt} = -\frac{1}{\tau} x(t) \qquad (11)$$

The solution of the above equation is the normalized exponential Debye-relaxation function (i.e. $x(t) = e^{-\frac{t}{\tau}}$), with relaxation time $\tau$. However, there are some experimental evidences that relaxation in several complex disordered systems deviates from the classical exponential Debye pattern [12-24]. Nowadays, it has proved that the fractional relaxation equation can be a successful mathematical construct that reflects the main features of evolution of such systems. The commonest way to obtain a fractional differential equation for describing the evolution of a typical system is to generalize the ordinary derivative in the standard differential equation into the fractional derivative

$$\frac{d}{dt} \rightarrow \frac{1}{\eta^{1-\alpha}} \frac{d^\alpha}{dt^\alpha} \qquad (12)$$

where $\frac{d^\alpha}{dt^\alpha}$ denotes the Caputo's derivative operator of order $\alpha$ and, $\eta$ is a new parameter representing the fractional time components in the system[32] and its dimension is the second. In the case $\alpha = 1$ the expression transforms into ordinary time derivative operator

$$\frac{1}{\eta^{1-\alpha}} \frac{d^\alpha}{dt^\alpha}\bigg|_{\alpha=1} = \frac{d}{dt} \qquad (13)$$

Therefore we can easily arrive at the fractional relaxation equation by changing the first order derivative in the Eq. (10) to a derivative of an arbitrary order:

$$\frac{\tau}{\eta^{1-\alpha}} \frac{d^{\alpha} x(t)}{dt^{\alpha}} + x(t) = 0 \qquad 0 < \alpha \leq 1 \qquad (14)$$

with the solution:

$$x(t) = x(0) E_{\alpha}(-\eta^{1-\alpha}(\frac{t^{\alpha}}{\tau})) \qquad (15)$$

It is showed that this solution and this model for the relaxation processes can be successfully adopted to interpret experimental data on relaxation in several complex disordered systems.

The second example is the simple harmonic oscillator. The harmonic oscillator, given by the well-known second order linear differential equation with constant coefficients

$$m \frac{d^2 x}{dt^2} + kx = 0 \qquad (16)$$

is a cornerstone of classical mechanics [7]. We can obtain the differential equation of a simple fractional oscillator [25-31] by changing the second derivative in the harmonic oscillator equation to a derivative of an arbitrary order (Eq. (12)):

$$\frac{m}{\eta^{2(1-\alpha)}} \frac{d^{2\alpha} x}{dt^{2\alpha}} + kx = 0 \qquad 0 < \alpha \leq 1 \qquad (17)$$

We can write the above equation also as

$$\frac{d^{2\alpha} x}{dt^{2\alpha}} + \eta^{2(1-\alpha)} \frac{k}{m} x = 0 \qquad (18)$$

where the parameter $\omega_f$ defined by

$$\omega_f^2 = \eta^{2(1-\alpha)} \omega^2 \qquad (19)$$

and $\omega^2 = \frac{k}{m}$, so we can rewrite the fractional differential equation of the system as

$$\frac{d^{2\alpha} x}{dt^{2\alpha}} + \omega_f^2 x = 0 \qquad (20)$$

The solution of the above equation reads:

$$x(t) = x(0) E_{2\alpha}(-\omega_f^2 t^{2\alpha}) + \dot{x}(0) t E_{2\alpha,2}(-\omega_f^2 t^{2\alpha}) \qquad (21)$$

Now if we choose $x(0) = 1$ and $\dot{x}(0) = 0$ as the initial condition, the solution becomes

$$x(t) = E_{2\alpha}(-\omega_f^2 t^{2\alpha}) \qquad (22)$$

We can easily see that as $\alpha \to 1$, above equations gives

$$E_{2\alpha}(-\omega_f^2 t^{2\alpha}) = E_2(-\omega^2 t^2) = \cosh(\sqrt{-\omega^2 t^2}) = \cosh(i\omega t) = \cos(\omega t) \qquad (23)$$

As we can see from Eq. (22), the displacement of the fractional oscillator is essentially described by the Mittag–Leffler function $E_{2\alpha}(-\omega_f^2 t^{2\alpha})$ for our considered initial conditions. It is showed by numerical calculations that the displacement of the fractional oscillator varies as a function of time and how this time variation depends on the parameter $\alpha$ [25]. Also it is proved that, if $\alpha$ is less than 1 the displacement shows the behavior of a damped harmonic oscillator. As a result, in consistent with the case of simple harmonic oscillator, the total energy of simple fractional oscillator will not be a constant. What is surprising is that the damping of fractional oscillator is intrinsic to the equation of motion and not introduced by additional forces as in the case of a damped harmonic oscillator. Up to now, the source of this intrinsic damping is not clearly understood. However, there are some attempts in this regard. For example an interesting formulation of the notion of intrinsic damping force has been proposed in Refs. [29, 30].

## 4- Classical electromagnetism: A plane wave with time decaying amplitude

In classical electromagnetism, behavior of electric fields ($\vec{E}$), magnetic fields ($\vec{B}$) and their relation to their sources charge density ($\rho(\vec{r},t)$), and current density ($\vec{j}(\vec{r},t)$), is described by the following Maxwell's equations:

$$\vec{\nabla}.\vec{E} = \frac{4\pi}{\varepsilon}\rho(\vec{r},t) \tag{24}$$

$$\vec{\nabla}.\vec{B} = 0 \tag{25}$$

$$\vec{\nabla}\times\vec{E} = -\frac{1}{c}\frac{\partial\vec{B}}{\partial t} \tag{26}$$

$$\vec{\nabla}\times\vec{B} = \frac{4\pi\mu}{c}\vec{j}(\vec{r},t) + \frac{\varepsilon\mu}{c}\frac{\partial\vec{E}}{\partial t} \tag{27}$$

Where $\varepsilon$ and $\mu$ are electric permittivity and magnetic permeability, respectively. Now, introducing the potentials, vector $\vec{A}(x_i,t)$ and scalar $\varphi(x_i,t)$

$$\vec{B} = \vec{\nabla}\times\vec{A} \tag{28}$$

$$\vec{E} = -\frac{1}{c}\frac{\partial\vec{A}}{\partial t} - \vec{\nabla}\varphi \tag{29}$$

and using the Lorenz gauge condition we obtain the following decoupled differential equations for the potentials

$$\Delta\vec{A}(\vec{r},t) - \frac{\varepsilon\mu}{c^2}\frac{\partial^2\vec{A}(\vec{r},t)}{\partial t^2} = -\frac{4\pi}{c}\vec{j}(\vec{r},t) \tag{30}$$

$$\Delta\varphi(\vec{r},t) - \frac{\varepsilon\mu}{c^2}\frac{\partial^2\varphi(\vec{r},t)}{\partial t^2} = -\frac{4\pi}{\varepsilon}\rho(\vec{r},t) \tag{31}$$

where $\frac{\varepsilon\mu}{c^2} = \frac{1}{v^2}$. $v$ is the velocity of the light in the medium. Furthermore, for a particle with charge q in the presence of electric and magnetic field we can write the Lorentz force as

$$\vec{F}_L = q(\vec{E} + \vec{v}\times\vec{B}) \tag{32}$$

In terms of scalar and vector potentials, Eq. (28, 29), we may write the Lorentz force as

$$\vec{F}_L = q(-\frac{1}{c}\frac{\partial\vec{A}}{\partial t} - \vec{\nabla}\varphi + \vec{v}\times(\vec{\nabla}\times\vec{A})) \tag{33}$$

As we saw in previous section, in classical mechanics, the fractional formalism leads to relaxation and oscillation processes that exhibit memory and delay. This fractional nonlocal formalism is also applicable on materials and media that have electromagnetic memory properties. So the generalized fractional Maxwell's equations can give us new models that can be used in these complex systems. Up to now, several different versions of fractional electromagnetism based on the different approaches to fractional vector calculus have been investigated [33-38]. However, in this paper we study a new approach on this area [32]. The idea is in fact, to write the ordinary differential wave equations in the fractional form with respect to $t$.

$$\vec{\nabla}.\vec{E} = \frac{4\pi}{\varepsilon}\rho(\vec{r},t) \tag{34}$$

$$\vec{\nabla}.\vec{B} = 0 \tag{35}$$

$$\vec{\nabla}\times\vec{E} = -\frac{1}{c}\frac{1}{\eta^{1-\alpha}}\frac{\partial^\alpha\vec{B}}{\partial t^\alpha} \qquad 0<\alpha\leq 1 \tag{36}$$

$$\vec{\nabla}\times\vec{B} = \frac{4\pi\mu}{c}\vec{j}(\vec{r},t) + \frac{\varepsilon\mu}{c}\frac{1}{\eta^{1-\alpha}}\frac{\partial^\alpha\vec{E}}{\partial t^\alpha} \qquad 0<\alpha\leq 1 \tag{37}$$

And the Eq. (28, 29) become

$$\vec{B} = \vec{\nabla} \times \vec{A} \tag{38}$$

$$\vec{E} = -\frac{1}{c\eta^{1-\alpha}} \frac{\partial^\alpha \vec{A}}{\partial^\alpha t} - \vec{\nabla}\varphi \qquad 0 < \alpha \le 1 \tag{39}$$

And the Lorentz force Eq. (33) becomes

$$\vec{F}_L = q(-\frac{1}{c\eta^{1-\alpha}} \frac{\partial^\alpha \vec{A}}{\partial^\alpha t} - \vec{\nabla}\varphi + \vec{v} \times (\vec{\nabla} \times \vec{A})) \qquad 0 < \alpha \le 1 \tag{40}$$

Then, applying the Lorentz gauge condition we obtain the corresponding time fractional wave equations for the potentials

$$\Delta \vec{A}(\vec{r},t) - \frac{\varepsilon\mu}{c^2} \frac{1}{\eta^{2(1-\alpha)}} \frac{\partial^{2\alpha} \vec{A}(\vec{r},t)}{\partial t^{2\alpha}} = -\frac{4\pi}{c} \vec{j}(\vec{r},t) \tag{41}$$

$$\Delta \varphi(\vec{r},t) - \frac{\varepsilon\mu}{c^2} \frac{1}{\eta^{2(1-\alpha)}} \frac{\partial^{2\alpha} \varphi(\vec{r},t)}{\partial t^{2\alpha}} = -\frac{4\pi}{\varepsilon} \rho(\vec{r},t) \tag{42}$$

If, $\rho = 0$, and, $\vec{j} = 0$, we have the homogeneous fractional differential equations

$$\Delta \vec{A}(\vec{r},t) - \frac{\varepsilon\mu}{c^2} \frac{1}{\eta^{2(1-\alpha)}} \frac{\partial^{2\alpha} \vec{A}(\vec{r},t)}{\partial t^{2\alpha}} = 0 \tag{43}$$

$$\Delta \varphi(\vec{r},t) - \frac{\varepsilon\mu}{c^2} \frac{1}{\eta^{2(1-\alpha)}} \frac{\partial^{2\alpha} \varphi(\vec{r},t)}{\partial t^{2\alpha}} = 0 \tag{44}$$

We are interested in the analysis of the electromagnetic fields in the medium starting from the equations. Now, we can write the fractional equations in the following compact form

$$\frac{\partial^2 z(x,t)}{\partial x^2} - \frac{\varepsilon\mu}{c^2} \frac{1}{\eta^{2(1-\alpha)}} \frac{\partial^{2\alpha} z(x,t)}{\partial t^{2\alpha}} = 0 \tag{45}$$

where $z(x,t)$ represents both $\vec{A}(\vec{r},t)$ and $\varphi(\vec{r},t)$. We consider a polarized electromagnetic wave, then $A_x = 0, A_y \ne 0, A_z \ne 0$. A particular solution of this equation may be found in the form

$$z(x,t) = z_0 e^{-ikx} u(t) \tag{46}$$

where $k$ is the wave vector in the $x$ direction and $z_0$ is a constant. Substituting into Eq. (45) we obtain

$$\frac{d^{2\alpha} u(t)}{dt^{2\alpha}} + \Omega_f^2 u(t) = 0 \tag{47}$$

Where

$$\Omega_f^2 = v^2 k^2 \eta^{2(1-\alpha)} = \Omega^2 \eta^{2(1-\alpha)} \tag{48}$$

and $\Omega$ is the fundamental frequency of the electromagnetic wave. The solution of this equation may be

$$u(t) = E_{2\alpha}(-\Omega_f^2 t^{2\alpha}) \tag{49}$$

Substituting this expression in Eq. (46) we have a particular solution of the equation as

$$z(x,t) = z_0 e^{-ikx} E_{2\alpha}(-\Omega_f^2 t^{2\alpha}) \tag{50}$$

We can easily see that in the case $\alpha = 1$, the solution to the equation is

$$z(x,t) = \text{Re}(z_0 e^{i(\Omega t - kx)}) \tag{51}$$

which defines a periodic, with fundamental period $T = 2\pi\Omega$, monochromatic wave in the, $x$, direction and in time, $t$. This result is very well known from the ordinary electromagnetic waves theory.
However for the arbitrary case of $\alpha$ ($0 < \alpha < 1$) the solution is periodic only respect to x and it is not periodic with respect to $t$. The solution represents a plane wave with time decaying amplitude.

For example for the case $\alpha = \frac{1}{2}$ we have

$$u(t) = E_1(-\eta\Omega^2 t) = e^{-\eta\Omega^2 t} \tag{52}$$

Therefore the solution is
$$z(x,t) = (z_0 e^{-\eta\Omega^2 t}) e^{-ikx} \tag{53}$$
Then, for this case the solution is periodic only respect to $x$ and it is not periodic with respect to $t$. In fact the solution represents a plane wave with time decaying amplitude.

## 5- Quantum mechanics: Time fractional Schrödinger and Pauli equation

Nowadays, application of the fractional calculus to quantum processes is a new and fast developing part of quantum physics which studies nonlocal quantum phenomena. Nonlocal effects may occur in space and time. In the time domain the extension from a local to a nonlocal description becomes manifest as a memory effect. Therefore the underlying fundamental processes become of non-Markovian type .In the realm of non-relativistic quantum mechanics [41, 42], Schrödinger equation represents a fundamental equation to study many quantum processes

$$-\frac{\hbar^2}{2m}\frac{\partial^2 \Psi}{\partial x^2} + V(x)\Psi = i\hbar\frac{\partial \Psi}{\partial t} \tag{54}$$

Recently the time fractional Schrödinger equation, which has a Caputo fractional time derivative, was considered by Naber [46], in order to describe non-Markovian evolution in quantum mechanics. The general idea to obtain the time fractional Schrödinger equation is to keep the position and momentum operators in the usual form and replacing $i\hbar\frac{\partial}{\partial t} \to (i\hbar_\alpha)\,{}_0^c D_t^\alpha$ or $i\hbar\frac{\partial}{\partial t} \to (i^\alpha \hbar_\alpha)\,{}_0^c D_t^\alpha$, where ${}_0^C D_t^\alpha$ denotes the Caputo's derivative operator of order $\alpha$ and $\hbar_\alpha = M_P c^2 T_P^\alpha$ is a scaled Planck constant. Also the parameters $M_P$ and $T_P$ are Planck mass and Planck time, respectively, are defined as

$$T_P = \sqrt{\frac{G\hbar}{c^5}} \quad , \quad M_P = \sqrt{\frac{c\hbar}{G}} \tag{55}$$

where G and c are the gravitational constant and the speed of light in vacuum, respectively.
Naber gives some arguments in favour of the latter case and many authors follow him therein [47, 48].However one can consider the former one as a possible case for studying time fractional Schrödinger equation. For instance the wave function and the probability density for a free particle within this type of time fractional Schrödinger equation

$$-\frac{\hbar^2}{2m}\frac{\partial^2 \Psi}{\partial x^2} + V(x)\Psi = (i\hbar_\alpha)\,{}_0^c D_t^\alpha \Psi \qquad 0 < \alpha \leq 1 \tag{56}$$

have been studied in Ref. [49].
As we mentioned above, one can introduce the time fractional Schrödinger equation to describe non-Markovian evolution in quantum realm. Now we generalize the time fractional Schrödinger equation Eq. (56) and obtain the following time fractional Pauli equation [50],

$$[\frac{1}{2m}(\hat{P} - \frac{e}{c}A)^2 + e\phi + \mu_B \hat{\sigma}.B]\Psi = (i\hbar_\alpha)\,{}_0^c D_t^\alpha \Psi \qquad 0 < \alpha \leq 1 \tag{57}$$

One can use this equation to discuss the electron spin precession problem in a homogeneous constant magnetic field [50]. Here we assume that, the electron is fixed at a certain location and its spin is the only degree of freedom. Also, let the magnetic field consist of a constant field $\vec{B}$ in the Z direction (i.e. $\vec{B} = B_0 \hat{k}$) .Therefore, that part of the time fractional Pauli equation Eq. (57), which contains the spin yields

$$(i\hbar_\alpha)\,{}_0^c D_t^\alpha \chi = \hbar_\alpha \omega_L^\alpha \hat{\sigma}_z \chi \tag{58}$$

Where $\omega_L = -\dfrac{eB}{2mc}$ are the so-called Larmor frequency and $\hat{\sigma}$ is the well-known Pauli matrices for a spin $\dfrac{1}{2}$ particles. Science the Hamiltonian of our system is a $2 \times 2$ matrix, the spin function in arbitrary time (t) must be written as a column matrix of two components and can be derived as below,

$$\chi_\alpha(t) = \begin{pmatrix} a(t) \\ b(t) \end{pmatrix} = \begin{pmatrix} e^{i\gamma} \cos\left(\frac{\theta}{2}\right) E_\alpha(-i(\omega_L t)^\alpha) \\ e^{i\delta} \sin\left(\frac{\theta}{2}\right) E_\alpha(i(\omega_L t)^\alpha) \end{pmatrix} \quad (59)$$

Where $\gamma$ and $\delta$ are arbitrary phase constants.

Now, by use of Eq. (59) we able to calculate the probability for spin-up, $P_{\alpha\uparrow}$, and spin-down, $P_{\alpha\downarrow}$, at $t > 0$:

$$P_{\alpha\uparrow} = |a(t)|^2 = \cos^2\left(\frac{\theta}{2}\right)[E_\alpha(-i(\omega_L t)^\alpha) E_\alpha(i(\omega_L t)^\alpha)] \quad (60)$$

$$P_{\alpha\downarrow} = |b(t)|^2 = \sin^2\left(\frac{\theta}{2}\right)[E_\alpha(-i(\omega_L t)^\alpha) E_\alpha(i(\omega_L t)^\alpha)]. \quad (61)$$

We can explicitly see that as $\alpha \to 1$, above equations gives $P_{\alpha=1_{tot}} = P_{\alpha=1\uparrow} + P_{\alpha=1\downarrow} = 1$.
But for the arbitrary case of $\alpha$ ($0 < \alpha < 1$), we have
$$P_{\alpha_{tot}} = P_{\alpha\uparrow} + P_{\alpha\downarrow} = \cos_\alpha^2((\omega_L t)^\alpha) + \sin_\alpha^2((\omega_L t)^\alpha) \quad (62)$$
Where is obtained in terms of the sine and cosine ML functions Eq. (7). It is clearly seen that the total probability of upness and downness of electron's spin varies as a function of time and also it depends on the parameter $\alpha$.

**6- Summary and discussion**

Fractional calculus is a very useful tool in describing the evolution of systems with memory, which typically are dissipative and to complex systems. In recent decades it has attracted interest of researches in several areas of science. Specially, in the field of physics applications of fractional calculus have gained considerable popularity [3, 4] (and the references therein). In spite of these various applications, there are some important challenges. For example physical interpretation for the fractional derivative is not completely clarified yet.
In this review, we present some fundamental classical and quantum phenomena in the framework of time fractional formalism in order to provide a deeper understanding of the physical interpretations of fractional derivative. We have seen that, a simple fractional oscillator behaves like a damped harmonic oscillator. What is surprising is that the damping is intrinsic to the equation of motion and not introduced by additional forces as in the case of a damped harmonic oscillator. Also, in the case of fractional electromagnetism we see that behavior of electromagnetic waves is not same as the standard ones. In fact we see that the fractional Maxwell's equations lead to the plan wave with time decaying amplitude (Eq. (50, 53)). It is showed that amplitude of this plane wave varies as a function of time and this time variation depends explicitly on the parameter $\alpha$ (the order of the fractional derivative)). Finally we see that total probability of upness and downness of electron's spin Eq. (62) is not equal to unity and it depends on $t$ and the parameter $\alpha$, as well. The interpretation of this time dependent probability is an open area of research. It is worth noticing that an expansion method has been proposed [28, 30] to discuss the dynamics in the media where the order of the fractional derivative is close to an integer number. It will be of interest to consider above mentioned phenomena within this scheme. We hope to report on these issues in the future.

# References

[1] I. Podlubny, Fractional Differential Equations (Academic Press, New York, 1999).
[2] R. Hilfer, Applications of Fractional Calculus in Physics (World Scientific, Singapore, 2000).
[3] R. Herrmann, Fractional Calculus (World Scientific Press, 2011).
[4] V.E. Tarasov, Fractional Dynamics (Springer, HEP, 2011).
[5] G.M.Mittag-Lefffler, C. R. Acad. Sci. Paris (Ser. II) **136** (1902) 937-939.
[6] J. Peng, K. Li, J. Math. Anal. Appl. **370** (2010) 635–638
[7] L. Landau and E. Lifshitz, Mechanics, 3rd ed. (Butterworth-Heinemann, Boston, 1976), Vol. 1.
[8] F. Riewe, Phys. Rev. E **53** (1996) 1890.
[9] F. Riewe, Phys. Rev. E **55** (1997) 3581.
[10] A. Ebaid, Applied Mathematical Modelling **35** (2011) 1231–1239.
[11] K. S. Fa, Physica A **350** (2005) 199–206.
[12] W.G. Glöckle and T.F. Nonnenmacher, Biophys. J. **68** (1995) 46.
[13] M.F. Shlesinger, G.M. Zazlavsky, J. Klafter, Nature **363** (1993) 31–37.
[14] R. Metzler, J. Klafter, Phys. Rep. **339** (2000) 1–77.
[15] R. Metzler and J. Klafter, J. Non-Cryst. Solids **305** (2002) 81.
[16] R. Metzler and J. Klafter, Biophys. J. **85** (2003) 2776.
[17] Y. Feldman, A. Puzenko, Y. Ryabov, Chem. Phys. **284** (2002) 139–168.
[18] R. Hilfer, Fractals **11** (2003) 251–257.
[19] R. Hilfer, Chem. Phys. **284** (2002) 399–408.
[20] R. Hilfer and L. Anton, Phys. Rev. E **51** (1995) R848.
[21] K. Weron and A. Klauzer, Ferroelectrics **236** (2000) 59.
[22] R. K. Saxena, A.M. Mathai and H.J. Haubold, Astrophys. Space Sci. **282** (2002) 281.
[23] M.N. Berberan-Santos, J. Math. Chem. **38** (2005) 165.
[24] F. Mainardi, J. Alloys Comp., **211/212** (1994) 534–538.
[25] B. N. Narahari Achar, J. W. Hanneken, T. Enck, and T. Clarke,Physica A **287**, (2001) 361–367.
[26] B. N. Narahari Achar, J. W. Hanneken, and T. Clarke, Physica A **309**, (2002) 275 – 288.
[27] B. N. Narahari Achar, J. W. Hanneken, T. Clarke, Physica A **339** (2004) 311 – 319.
[28] V. E. Tarasov, G.M. Zaslavsky, Physica A **368**(2006) 399.
[29] A. Tofghi, Physica A **329** (2003) 29 – 34.
[30] A. Tofighi, H .Nasrolahpour, Physica A **374**(2007) 41.
[31] A. A. Stanislavsky,Phys. Rev. E ,**70** (2004) 051103
[32] J. F. Gomez, J.J. Rosales, J.J. Bernal, V.I. Tkach, M. Guia, Eprint: math-ph / 1108.6292.
[33] M. J. Lazo, Eprint: math-ph / 1108.3493.
[34] V. E. Tarasov, Ann. Phys. **323** (2008) 2756–2778.
[35] Q. A. Naqvi, M. Abbas, Opt. Commun. **241** (2004) 349-355.
[36] A. Hussain, Q.A. Naqvi, Prog. Electromagn. Res. **59** (2006) 199-213;
[37] A. Hussain, S. Ishfaq, Q.A. Naqvi, Prog. Electromagn. Res. **63** (2006) 319-335.
[38] N. Engheta, Microwave and Opt. Technol. Lett. **17 (2)** (1998) 86-91.
[39] V. E. Tarasov, Int. J. Theor. Phys. **49**(2010)293-303.
[40] H. Nasrolahpour, Prespacetime J., **2 (8)** (2011)1264-1269.
[41] J. J. Sakurai, Modern Quantum Mechanics (Addison-Wesley, New York, 1994).
[42] W. Greiner, Quantum Mechanics: An Introduction, 4th ed. (Springer, Berlin, 2001).
[43] N. Laskin, Phys.Lett. A **268** (2000) 298.
[44] N. Laskin, Phys. Rev. E **62** (2000) 3135.
[45] N. Laskin, Phys. Rev. E **66** (2002) 056108
[46] M. Naber, J. Math. Phys. **45** (2004) 3339.
[47] A. Tofighi, Acta Phys. Pol. A. Vol. **116** (2009)114-118.
[48] H. Ertik, D. Demirhan, H. Şirin, and F. Büyükkılıç, J. Math. Phys. **51**(2010) 082102.
[49] M. Bhatti, Int. J. Contemp. Math. Sci. **2** (2007) 943.
[50] H. Nasrolahpour, Prespacetime J., **2 (13)** (2011) 2053-2059.
[51] A. Raspini, Phys. Scr.**64** (2001)20.
[52] R. Herrmann, Phys. Lett. A **372** (2008) 5515.
[53] S. I. Muslih, Om P. Agrawal, D. Baleanu, J. Phys. A: Math. Theor. **43** (2010) 055203.

[54] E. Goldfain, Chaos, Solitons & Fractals **28** (2006) 913–922.
[55] E. Goldfain, Comm. Non. Sci. Num. Siml. , **13**(2008) 1397-1404.
[56] E. Goldfain, Comm. in Nonlin. Dynamics and Numer. Simulation, **14** (2009) 1431-1438.
[57] R. A. El-Nabulsi , Chaos, Solitons & Fractals **42** (2009) 2614–2622.
[58] R. A. El-Nabulsi, Chaos, Solitons & Fractals **41** (2009) 2262–2270.